\documentstyle[12pt,color,epsfig,subfigure]{article}
\def\baselinestretch{1.3}
\advance\textheight by \topskip
\newcommand{\comment}[1]{}

\begin{document}
\tolerance=100000
\thispagestyle{empty}
\setcounter{page}{0}
\topmargin -0.1in
\headsep 30pt
\footskip 40pt
\oddsidemargin 12pt
\evensidemargin -16pt
\textheight 8.5in
\textwidth 6.5in
\parindent 20pt
 
\def\baselinestretch{1.5}
\newcommand{\newc}{\newcommand}
\def\preprint{{preprint}}
\def\Ord{\lower .7ex\hbox{$\;\stackrel{\textstyle <}{\sim}\;$}}
\def\OOrd{\lower .7ex\hbox{$\;\stackrel{\textstyle >}{\sim}\;$}}
\def\cO#1{{\cal{O}}\left(#1\right)}
\newc{\order}{{\cal O}}
\def\lag             {{\cal L}}
\def\Lag             {{\cal L}}
\def\lum             {{\cal L}}
\def\R               {{\cal R}}
\def\Rsq             {{\cal R}^{\sq}}
\def\Rst             {{\cal R}^{\st}}
\def\Rsb             {{\cal R}^{\sb}}
\def\M               {{\cal M}}
\def\Oas             {{\cal O}(\alpha_{s})}
\def\Vcal            {{\cal V}}
\def\Wcal            {{\cal W}}
\newc{\be}{\begin{equation}}
\newc{\ee}{\end{equation}}
\newc{\br}{\begin{eqnarray}}
\newc{\er}{\end{eqnarray}}
\newc{\ba}{\begin{array}}
\newc{\ea}{\end{array}}
\newc{\bi}{\begin{itemize}}
\newc{\ei}{\end{itemize}}
\newc{\bn}{\begin{enumerate}}
\newc{\en}{\end{enumerate}}
\newc{\bc}{\begin{center}}
\newc{\ec}{\end{center}}
\newc{\ul}{\underline}
\newc{\ol}{\overline}
\newc{\ra}{\rightarrow}
\newc{\lra}{\longrightarrow}
\newc{\wt}{\widetilde}
\newc{\til}{\tilde}
\def\kr              {^{\dagger}}
\newc{\wh}{\widehat}
\newc{\ti}{\times}
\newc{\Dir}{\kern -6.4pt\Big{/}}
\newc{\Dirin}{\kern -10.4pt\Big{/}\kern 4.4pt}
\newc{\DDir}{\kern -10.6pt\Big{/}}
\newc{\DGir}{\kern -6.0pt\Big{/}}
\newc{\sig}{\sigma}
\newc{\sigmalstop}{\sig_{\lstoppair}}
\newc{\Sig}{\Sigma}  
\newc{\del}{\delta}
\newc{\Del}{\Delta}
\newc{\lam}{\lambda}
\newc{\Lam}{\Lambda}
\newc{\gam}{\gamma}
\newc{\Gam}{\Gamma}
\newc{\eps}{\epsilon}
\newc{\Eps}{\Epsilon}
\newc{\kap}{\kappa}
\newc{\Kap}{\Kappa}
\newc{\modulus}[1]{\left| #1 \right|}
\newc{\eq}[1]{(\ref{eq:#1})}
\newc{\eqs}[2]{(\ref{eq:#1},\ref{eq:#2})}
\newc{\etal}{{\it et al.}\ }
\newc{\ibid}{{\it ibid}.}
\newc{\ibidem}{{\it ibidem}.}
\newc{\eg}{{\it e.g.}\ }
\newc{\ie}{{\it i.e.}\ }
\def \viz{\emph{viz.}}
\def \etc{\emph{etc. }}
\newc{\nonum}{\nonumber}
\newc{\lab}[1]{\label{eq:#1}}
\newc{\dpr}[2]{({#1}\cdot{#2})}
\newc{\lt}{\stackrel{<}}
\newc{\gt}{\stackrel{>}}
\newc{\lsimeq}{\stackrel{<}{\sim}}
\newc{\gsimeq}{\stackrel{>}{\sim}}
\def\lsim{\buildrel{\scriptscriptstyle <}\over{\scriptscriptstyle\sim}}
\def\gsim{\buildrel{\scriptscriptstyle >}\over{\scriptscriptstyle\sim}}
\def\lapp{\mathrel{\rlap{\raise.5ex\hbox{$<$}}
                    {\lower.5ex\hbox{$\sim$}}}}
\def\gapp{\mathrel{\rlap{\raise.5ex\hbox{$>$}}
                    {\lower.5ex\hbox{$\sim$}}}}
\newc{\half}{\frac{1}{2}}
\newcommand {\nnc}        {{\overline{\mathrm N}_{95}}}
\newcommand {\dm}         {\Delta m}
\newcommand {\dM}         {\Delta M}
\def\bra{\langle}
\def\ket{\rangle}
\def\cO#1{{\cal{O}}\left(#1\right)}
\def \DM{{\Delta{m}}}
\newc{\bQ}{\ol{Q}}
\newc{\dota}{\dot{\alpha }}
\newc{\dotb}{\dot{\beta }}
\newc{\dotd}{\dot{\delta }}
\newc{\nindnt}{\noindent}

\newcommand{\medf}[2] {{\footnotesize{\frac{#1}{#2}} }}
\newcommand{\smaf}[2] {{\textstyle \frac{#1}{#2} }}
\def\onesq            {{\textstyle \frac{1}{\sqrt{2}} }}
\def\onehf            {{\textstyle \frac{1}{2} }}
\def\oneth            {{\textstyle \frac{1}{3} }}
\def\twoth            {{\textstyle \frac{2}{3} }}
\def\onefo            {{\textstyle \frac{1}{4} }}
\def\forth            {{\textstyle \frac{4}{3} }}

\newc{\matth}{\mathsurround=0pt}
\def\ML{\ifmmode{{\mathaccent"7E M}_L}
             \else{${\mathaccent"7E M}_L$}\fi}
\def\MR{\ifmmode{{\mathaccent"7E M}_R}
             \else{${\mathaccent"7E M}_R$}\fi}
\newcommand{\s}{\\ \vspace*{-3mm} }

\def \ud { {1 \over 2} }
\def \ut { {1 \over 3} }
\def \td { {3 \over 2} }
\newc{\mr}{\mathrm}
\def\dh {\partial }
\def \cs { cross-section }
\def \css { cross-sections }
\def \cm { centre of mass }
\def \cms { centre of mass energy }
\def \cc { coupling constant }
\def \ccs {coupling constants }
\def \gc {gauge coupling }
\def \gcc {gauge coupling constant }
\def \gccs {gauge coupling constants }
\def \yc {Yukawa coupling }
\def \ycc {Yukawa coupling constant }
\def \pp {{parameter }}
\def \pps {{parameters }} 
\def \ps {parameter space }
\def \pss {parameter spaces }
\def \vv {vice versa }

\newc{\siminf}{\mbox{$_{\sim}$ {\small {\hspace{-1.em}{$<$}}}    }}
\newc{\simsup}{\mbox{$_{\sim}$ {\small {\hspace{-1.em}{$>$}}}    }}


\newc {\Zboson}{{\mathrm Z}^{0}}
\newc{\thetaw}{\theta_W}
\newc{\mbot}{{m_b}}
\newc{\mtop}{{m_t}}
\newc{\sm}{${\cal {SM}}$}
\newc{\as}{\alpha_s}
\newc{\aem}{\alpha_{em}}
\def \PI{{\pi^{\pm}}}
\newc{\ppbar}{\mbox{$p\ol{p}$}}
\newc{\bbbar}{\mbox{$b\ol{b}$}}
\newc{\ccbar}{\mbox{$c\ol{c}$}}
\newc{\ttbar}{\mbox{$t\ol{t}$}}
\newc{\eebar}{\mbox{$e\ol{e}$}}
\newc{\zzero}{\mbox{$Z^0$}}
\def \gamz{\Gam_Z}
\newc{\wplus}{\mbox{$W^+$}}
\newc{\wminus}{\mbox{$W^-$}}
\newc{\ellp}{\ell^+}
\newc{\ellm}{\ell^-}
\newc{\elp}{\mbox{$e^+$}}
\newc{\elm}{\mbox{$e^-$}}
\newc{\elpm}{\mbox{$e^{\pm}$}}
\newc{\qbar}     {\mbox{$\ol{q}$}}
\def \ewgroup{SU(2)_L \otimes U(1)_Y}
\def \smgroup{SU(3)_C \otimes SU(2)_L \otimes U(1)_Y}
\def \smcolorem{SU(3)_C \otimes U(1)_{em}}

\def \SSM  {Supersymmetric Standard Model}
\def \poincare{Poincare$\acute{e}$}
\def \superspace{\emph{superspace}}
\def \sfs{\emph{superfields}}
\def \superpot{\emph{superpotential}}
\def \csf{\emph{chiral superfield}}
\def \csfs{\emph{chiral superfields}}
\def \vsf{\emph{vector superfield }}
\def \vsfs{\emph{vector superfields}}
\newc{\Ebar}{{\bar E}}
\newc{\Dbar}{{\bar D}}
\newc{\Ubar}{{\bar U}}
\newc{\susy}{{{SUSY}}}
\newc{\msusy}{{{M_{SUSY}}}}

\def\photino{\ifmmode{\mathaccent"7E \gam}\else{$\mathaccent"7E \gam$}\fi}
\def\taugluino{\ifmmode{\tau_{\mathaccent"7E g}}
             \else{$\tau_{\mathaccent"7E g}$}\fi}
\def\mphotino{\ifmmode{m_{\mathaccent"7E \gam}}
             \else{$m_{\mathaccent"7E \gam}$}\fi}
\newc{\gl}   {\mbox{$\wt{g}$}}
\newc{\mgl}  {\mbox{$m_{\gl}$}}
\def \charginopm{{\wt\chi}^{\pm}}
\def \mcharginopm{m_{\charginopm}}
\def \mchpmmin {\mcharginopm^{min}}
\def \chonep {{\wt\chi_1^+}}
\def \chone {{\wt\chi_1}}
\def \ch2p {{\wt\chi_2^+}}
\def \chonem {{\wt\chi_1^-}}
\def \ch2m {{\wt\chi_2^-}}
\def \chplus {{\wt\chi^+}}
\def \chminus {{\wt\chi^-}}
\def \chonip{{\wt\chi_i}^{+}}
\def \chonim{{\wt\chi_i}^{-}}
\def \chonipm{{\wt\chi_i}^{\pm}}
\def \chonjp{{\wt\chi_j}^{+}}
\def \chonjm{{\wt\chi_j}^{-}}
\def \chonjpm{{\wt\chi_j}^{\pm}}
\def \chonepm{{\wt\chi_1}^{\pm}}
\def \chonemp{{\wt\chi_1}^{\mp}}
\def \mchonepm{m_{\chonepm}}
\def \mchonemp{m_{\chonemp}}
\def \chtwopm{{\wt\chi_2}^{\pm}}
\def \mchtwopm{m_{\chtwopm}}
\newc{\dmchi}{\Delta m_{\wt\chi}}


\def \vlsp{\emph{VLSP}}
\def \lspi{\wt\chi_i^0}
\def \mlspi{m_{\lspi}}
\def \lspj{\wt\chi_j^0}
\def \mlspj{m_{\lspj}}
\def \lspone{\wt\chi_1^0}
\def \mlspone{m_{\lspone}}
\def \lsptwo{\wt\chi_2^0}
\def \mlsptwo{m_{\lsptwo}}
\def \lspthree{\wt\chi_3^0}
\def \mlspthree{m_{\lspthree}}
\def \lspfour{\wt\chi_4^0}
\def \mlspfour{m_{\lspfour}}


\newc{\sele}{\wt{\mathrm e}}
\newc{\sell}{\wt{\ell}}
\def \msell{m_{\sell}}
\def \slepone{\wt\ell_1}
\def \mslepone{m_{\slepone}}
\def \smuone{\wt\mu_1}
\def \msmuone{m_{\smuone}}
\def \stauone{\wt\tau}
\def \mstauone{m_{\stauone}}
\def \snu{\wt{\nu}}
\def \snutau{\wt{\nu}_{\tau}}
\def \msnu{m_{\snu}}
\def \msnumu{m_{\snu_{\mu}}}
\def \barsnu{\wt{\bar{\nu}}}
\def \barsnul{\barsnu_{\ell}}
\def \snul{\snu_{\ell}}
\def \mbarsnu{m_{\barsnu}}
\newc{\snue}     {\mbox{$ \wt{\nu_e}$}}
\newc{\smu}{\wt{\mu}}
\newc{\stau}{\wt{\tau}}
\newc {\nuL} {\wt{\nu}_L}
\newc {\nuR} {\wt{\nu}_R}
\newc {\snub} {\bar{\wt{\nu}}}
\newc {\eL} {\wt{e}_L}
\newc {\eR} {\wt{e}_R}
\def \slep{\wt{l}}
\def \slepl{\wt{l}_L}
\def \mslepl{m_{\slepl}}
\def \slepr{\wt{l}_R}
\def \mslepr{m_{\slepr}}
\def \stau{\wt\tau}
\def \mstau{m_{\stau}}
\def \slepton{\wt\ell}
\def \mslepton{m_{\slepton}}
\def \mlhiggs{m_{h^0}}

\def \xr{X_{r}}

\def \sfer{\wt{f}}
\def \msfer{m_{\sfer}}
\def \sq{\wt{q}}
\def \msq{m_{\sq}}
\def \msquleft{m_{\tilde{u_L}}}
\def \msqurht{m_{\tilde{u_R}}}
\def \sql{\wt{q}_L}
\def \msql{m_{\sql}}
\def \sqr{\wt{q}_R}
\def \msqr{m_{\sqr}}
\newc{\msqot}  {\mbox{$m_(\sq_{1,2} )$}}
\newc{\sqbar}    {\mbox{$\bar{\wt{q}}$}}
\newc{\ssb}      {\mbox{$\squark\ol{\squark}$}}
\newc {\qL} {\wt{q}_L}
\newc {\qR} {\wt{q}_R}
\newc {\uL} {\wt{u}_L}
\newc {\uR} {\wt{u}_R}
\def \ul{\wt{u}_L}
\def \mul{m_{\ul}}
\newc {\dL} {\wt{d}_L}
\newc {\dR} {\wt{d}_R}
\newc {\cL} {\wt{c}_L}
\newc {\cR} {\wt{c}_R}
\newc {\sL} {\wt{s}_L}
\newc {\sR} {\wt{s}_R}
\newc {\tL} {\wt{t}_L}
\newc {\tR} {\wt{t}_R}
\newc {\stb} {\ol{\wt{t}}_1}
\newc {\sbot} {\wt{b}_1}
\newc {\msbot} {m_{\sbot}}
\newc {\sbotb} {\ol{\wt{b}}_1}
\newc {\bL} {\wt{b}_L}
\newc {\bR} {\wt{b}_R}
\def \mul{m_{\wt{u}_L}}
\def \mur{m_{\wt{u}_R}}
\def \mdl{m_{\wt{d}_L}}
\def \mdr{m_{\wt{d}_R}}
\def \mcl{m_{\wt{c}_L}}
\def \charml{\wt{c}_L}
\def \mcr{m_{\wt{c}_R}}
\newc{\csquark}  {\mbox{$\wt{c}$}}
\newc{\csquarkl} {\mbox{$\wt{c}_L$}}
\newc{\mcsl}     {\mbox{$m(\csquarkl)$}}
\def \msl{m_{\wt{s}_L}}
\def \msr{m_{\wt{s}_R}}
\def \mbl{m_{\wt{b}_L}}
\def \mbr{m_{\wt{b}_R}}
\def \mtl{m_{\wt{t}_L}}
\def \mtr{m_{\wt{t}_R}}
\def \st{\wt{t}}
\def \mst{m_{\st}}
\newc {\stopl}         {\wt{\mathrm{t}}_{\mathrm L}}
\newc {\stopr}         {\wt{\mathrm{t}}_{\mathrm R}}
\newc {\stoppair}      {\wt{\mathrm{t}}_{1}
\bar{\wt{\mathrm{t}}}_{1}}
\def \lstop{\wt{t}_{1}}
\def \lstopbar{\lstop^*}
\def \hstop{\wt{t}_{2}}
\def \hstopbar{\hstop^*}
\def \mlstop{m_{\lstop}}
\def \mhstop{m_{\hstop}}
\def \lstoppair{\lstop\lstop^*}
\def \hstoppair{\hstop\hstop^*}
\newc{\tsquark}  {\mbox{$\wt{t}$}}
\newc{\ttb}      {\mbox{$\tsquark\ol{\tsquark}$}}
\newc{\ttbone}   {\mbox{$\tsquark_1\ol{\tsquark}_1$}}
\def \tsq {top squark }
\def \tsqs {top squarks }
\def \tsql {ligtest top squark }
\def \tsqh {heaviest top squark }
\newc{\mix}{\theta_{\wt t}}
\newc{\cost}{\cos{\theta_{\wt t}}}
\newc{\sint}{\sin{\theta_{\wt t}}}
\newc{\costloop}{\cos{\theta_{\wt t_{loop}}}}
\def \lsbot{\wt{b}_{1}}
\def \lsbotbar{\lsbot^*}
\def \hsbot{\wt{b}_{2}}
\def \hsbotbar{\hsbot^*}
\def \mlsbot{m_{\lsbot}}
\def \mhsbot{m_{\hsbot}}
\def \lsbotpair{\lsbot\lsbot^*}
\def \hsbotpair{\hsbot\hsbot^*}
\newc{\mixsbot}{\theta_{\wt b}}

\def \mhone{m_{h_1}}
\def \hup{{H_u}}
\def \hdn{{H_d}}
\newc{\tb}{\tan\beta}
\newc{\cb}{\cot\beta}
\newc{\vev}[1]{{\left\langle #1\right\rangle}}

\def \abot{A_{b}}
\def \atop{A_{t}}
\def \atau{A_{\tau}}
\newc{\mhalf}{m_{1/2}}
\newc{\mzero} {\mbox{$m_0$}}
\newc{\azero} {\mbox{$A_0$}}

\newc{\lb}{\lam}
\newc{\lbp}{\lam^{\prime}}
\newc{\lbpp}{\lam^{\prime\prime}}
\newc{\rpv}{{\not \!\! R_p}}
\newc{\rpvm}{{\not  R_p}}
\newc{\rp}{R_{p}}
\newc{\rpmssm}{{RPC MSSM}}
\newc{\rpvmssm}{{RPV MSSM}}


\newc{\sbyb}{S/$\sqrt B$}
\newc{\pelp}{\mbox{$e^+$}}
\newc{\pelm}{\mbox{$e^-$}}
\newc{\pelpm}{\mbox{$e^{\pm}$}}
\newc{\epem}{\mbox{$e^+e^-$}}
\newc{\lplm}{\mbox{$\ell^+\ell^-$}}
\def \branch{\emph{BR}}
\def \branche{\branch(\lstop\ra be^{+}\nu_e \lspone)\ti \branch(\lstop^{*}\ra \bar{b}q\bar{q^{\prime}}\lspone)}
\def \branchmu{\branch(\lstop\ra b\mu^{+}\nu_{\mu} \lspone)\ti \branch(\lstop^{*}\ra \bar{b}q\bar{q^{\prime}}\lspone)}
\def\Ecm{\ifmmode{E_{\mathrm{cm}}}\else{$E_{\mathrm{cm}}$}\fi}
\newc{\rts}{\sqrt{s}}
\newc{\rtshat}{\sqrt{\hat s}}
\newc{\gev}{\,GeV}
\newc{\mev}{~{\rm MeV}}
\newc{\tev}  {\mbox{$\;{\rm TeV}$}}
\newc{\gevc} {\mbox{$\;{\rm GeV}/c$}}
\newc{\gevcc}{\mbox{$\;{\rm GeV}/c^2$}}
\newc{\intlum}{\mbox{${ \int {\cal L} \; dt}$}}
\newc{\call}{{\cal L}}
\def \met  {\mbox{${E\!\!\!\!/_T}$}}
\def \cpv  {\mbox{${CP\!\!\!\!/}$}}
\newc{\ptmiss}{/ \hskip-7pt p_T}
\def \eslash{\not \! E}
\def \etslash{\not \! E_T }
\def \ptslash{\not \! p_T }
\newc{\PT}{\mbox{$p_T$}}
\newc{\ET}{\mbox{$E_T$}}
\newc{\dedx}{\mbox{${\rm d}E/{\rm d}x$}}
\newc{\ifb}{\mbox{${\rm fb}^{-1}$}}
\newc{\ipb}{\mbox{${\rm pb}^{-1}$}}
\newc{\pb}{~{\rm pb}}
\newc{\fb}{~{\rm fb}}
\newc{\ycut}{y_{\mathrm{cut}}}
\newc{\chis}{\mbox{$\chi^{2}$}}
\def \hadron{\emph{hadron}}
\def \nlc{\emph{NLC }}
\def \lhc{\emph{LHC }}
\def \cdf{\emph{CDF }}
\def\dzero{\emptyset}
\def \tevatron{\emph{Tevatron }}
\def \lep{\emph{LEP }}
\def \jets{\emph{jets }}
\def \jet(s){\emph{jet(s) }}

\def\Crs{stroke [] 0 setdash exch hpt sub exch vpt add hpt2 vpt2 neg V currentpoint stroke 
hpt2 neg 0 R hpt2 vpt2 V stroke}
\def\loopdk{\lstop \ra c \lspone}
\def\brloopdk{\branch(\loopdk)}
\def\fourdk{\lstop \ra b \lspone  f \bar f'}
\def\brfourdk{\branch(\fourdk)}
\def\fourdklep{\lstop \ra b \lspone  \ell \nu_{\ell}}
\def\fourdkhad{\lstop \ra b \lspone  q \bar q'}
\def\brfourdklep{\branch(\fourdklep)}
\def\brfourdkhad{\branch(\fourdkhad)}
\def\twodk{\lstop \ra b \chonep}
\def\brtwodk{\branch(\twodk)}
\def\threedkslep{\lstop \ra b \wt{\ell} \nu_{\ell}}
\def\brthreedkslep{\branch(\threedkslep)}
\def\threedksnu{\lstop \ra b \wt{\nu_{\ell}} \ell }
\def\brthreedksnu{\branch(\threedksnu) }
\def\threedklsp{\lstop \ra b W \lspone }
\def\brthreedklsp{\\branch(\threedklsp) }
\def\topdk{t \ra \lstop \lspone}
\def\rpvdk{\lstop \ra e^+ d}
\def\brrpvdk{\branch(\rpvdk)}
\def\fonec{f_{11c}} 
\newc{\mpl}{M_{\rm Pl}}
\newc{\mgut}{M_{GUT}}
\newc{\mw}{M_{W}}
\newc{\mweak}{M_{weak}}
\newc{\mz}{M_{Z}}

\newc{\OPALColl}   {OPAL Collaboration}
\newc{\ALEPHColl}  {ALEPH Collaboration}
\newc{\DELPHIColl} {DELPHI Collaboration}
\newc{\XLColl}     {L3 Collaboration}
\newc{\JADEColl}   {JADE Collaboration}
\newc{\CDFColl}    {CDF Collaboration}
\newc{\DXColl}     {D0 Collaboration}
\newc{\HXColl}     {H1 Collaboration}
\newc{\ZEUSColl}   {ZEUS Collaboration}
\newc{\LEPColl}    {LEP Collaboration}
\newc{\ATLASColl}  {ATLAS Collaboration}
\newc{\CMSColl}    {CMS Collaboration}
\newc{\UAColl}    {UA Collaboration}
\newc{\KAMLANDColl}{KamLAND Collaboration}
\newc{\IMBColl}    {IMB Collaboration}
\newc{\KAMIOColl}  {Kamiokande Collaboration}
\newc{\SKAMIOColl} {Super-Kamiokande Collaboration}
\newc{\SUDANTColl} {Soudan-2 Collaboration}
\newc{\MACROColl}  {MACRO Collaboration}
\newc{\GALLEXColl} {GALLEX Collaboration}
\newc{\GNOColl}    {GNO Collaboration}
\newc{\SAGEColl}  {SAGE Collaboration}
\newc{\SNOColl}  {SNO Collaboration}
\newc{\CHOOZColl}  {CHOOZ Collaboration}
\newc{\PDGColl}  {Particle Data Group Collaboration}

\def\issue(#1,#2,#3){{\bf #1}, #2 (#3)}
\def\ASTR(#1,#2,#3){Astropart.\ Phys. \issue(#1,#2,#3)}
\def\AJ(#1,#2,#3){Astrophysical.\ Jour. \issue(#1,#2,#3)}
\def\AJS(#1,#2,#3){Astrophys.\ J.\ Suppl. \issue(#1,#2,#3)}
\def\APP(#1,#2,#3){Acta.\ Phys.\ Pol. \issue(#1,#2,#3)}
\def\JCAP(#1,#2,#3){Journal\ XX. \issue(#1,#2,#3)} 
\def\SC(#1,#2,#3){Science \issue(#1,#2,#3)}
\def\PRD(#1,#2,#3){Phys.\ Rev.\ D \issue(#1,#2,#3)}
\def\PR(#1,#2,#3){Phys.\ Rev.\ \issue(#1,#2,#3)} 
\def\PRC(#1,#2,#3){Phys.\ Rev.\ C \issue(#1,#2,#3)}
\def\NPB(#1,#2,#3){Nucl.\ Phys.\ B \issue(#1,#2,#3)}
\def\NPPS(#1,#2,#3){Nucl.\ Phys.\ Proc. \ Suppl \issue(#1,#2,#3)}
\def\NJP(#1,#2,#3){New.\ J.\ Phys. \issue(#1,#2,#3)}
\def\JP(#1,#2,#3){J.\ Phys.\issue(#1,#2,#3)}
\def\PL(#1,#2,#3){Phys.\ Lett. \issue(#1,#2,#3)}
\def\PLB(#1,#2,#3){Phys.\ Lett.\ B  \issue(#1,#2,#3)}
\def\ZP(#1,#2,#3){Z.\ Phys. \issue(#1,#2,#3)}
\def\ZPC(#1,#2,#3){Z.\ Phys.\ C  \issue(#1,#2,#3)}
\def\PREP(#1,#2,#3){Phys.\ Rep. \issue(#1,#2,#3)}
\def\PRL(#1,#2,#3){Phys.\ Rev.\ Lett. \issue(#1,#2,#3)}
\def\MPL(#1,#2,#3){Mod.\ Phys.\ Lett. \issue(#1,#2,#3)}
\def\RMP(#1,#2,#3){Rev.\ Mod.\ Phys. \issue(#1,#2,#3)}
\def\SJNP(#1,#2,#3){Sov.\ J.\ Nucl.\ Phys. \issue(#1,#2,#3)}
\def\CPC(#1,#2,#3){Comp.\ Phys.\ Comm. \issue(#1,#2,#3)}
\def\IJMPA(#1,#2,#3){Int.\ J.\ Mod. \ Phys.\ A \issue(#1,#2,#3)}
\def\MPLA(#1,#2,#3){Mod.\ Phys.\ Lett.\ A \issue(#1,#2,#3)}
\def\PTP(#1,#2,#3){Prog.\ Theor.\ Phys. \issue(#1,#2,#3)}
\def\RMP(#1,#2,#3){Rev.\ Mod.\ Phys. \issue(#1,#2,#3)}
\def\NIMA(#1,#2,#3){Nucl.\ Instrum.\ Methods \ A \issue(#1,#2,#3)}
\def\JHEP(#1,#2,#3){J.\ High\ Energy\ Phys. \issue(#1,#2,#3)}
\def\EPJC(#1,#2,#3){Eur.\ Phys.\ J.\ C \issue(#1,#2,#3)}
\def\RPP (#1,#2,#3){Rept.\ Prog.\ Phys. \issue(#1,#2,#3)}
\def\PPNP(#1,#2,#3){ Prog.\ Part.\ Nucl.\ Phys. \issue(#1,#2,#3)}
\newc{\PRDR}[3]{{Phys. Rev. D} {\bf #1}, Rapid  Communications, #2 (#3)}

\vspace{-1.5in}
\begin{flushright}
IISER/HEP/09/06
\end{flushright}
\begin{center}
{\Large \bf
Tracking down the elusive charginos / neutralinos 
through $\tau$ leptons at the Large Hadron Collider
}\\[1.00cm]
{\large Nabanita Bhattacharyya,\footnote{\it nabanita@iiserkol.ac.in}
        and Amitava Datta,\footnote{\it adatta@iiserkol.ac.in}}\\
 {\it  Indian Institute of Science Education and Research-Kolkata,
 P. O. Mohanpur, Pin: 741246, West Bengal, India.}\\
\end{center}
\vspace{.2cm}

\begin{abstract}

An unconstrained minimal supersymmetric standard model
with the superpartners of the strongly interacting
particles very heavy (close to the kinematic reach of the LHC
or even beyond it) and a relatively light electroweak sector is
considered. Using the event generator Pythia
it is shown that the 
1$\tau$-jet (tagged) + 2$l$ and 2$\tau$-jets (tagged) + 1$l$ 
signals with a reasonably hard $\etslash$ spectrum
either by themselves or in combination with the
conventional $3l$ signal, which is known to be of rather modest size
with a soft $\etslash$ spectrum, may
appreciably extend the reach of chargino-neutralino 
search at the LHC with 10 fb$^{-1}$ of integrated luminosity. 
This is especially so if the lighter
chargino and the second lightest neutralino decays via two body leptonic
modes with large BRs. The theoretical motivation of
this scenario, yielding large values of the fine-tuning parameters but
consistent with various indirect constraints including the dark matter relic
density, is briefly discussed. It is shown that in the minimal
supergravity (mSUGRA) model with an universal scalar mass at the GUT
scale,  the  signals involving
$\tau$-jets are not viable. Theoretically well-motivated
variations of these boundary conditions are, however, 
adequate for reviving these signals.

\end{abstract}
\newpage
\section{Introduction} 
\label{intro4}

Supersymmetry (SUSY) \cite{susy} 
is one of the most well motivated extensions 
of the standard model (SM) of particle physics. Moreover, it is
widely expected on general theoretical arguments like the
naturalness of the  Higgs boson 
mass \cite{natural1,natural2} 
that the masses of the superpartners of the SM particles,
collectively called the sparticles, should  be $\order$(1 TeV).
Therefore the  discovery of SUSY at the large hadron collider (LHC), the 
first accelerator designed for probing TeV scale physics, is
eagerly awaited.

Unfortunately the SUSY breaking mechanism and, consequently, the 
mass spectrum of the sparticles at the energies of experimental 
interest is still unknown. Thus one cannot apriorily exclude the 
possibility that a sub-set of the sparticles may very well be too  heavy 
and escape detection at the LHC. Whether the relatively light sparticles are
adequate to produce observable signatures at the LHC is then one of the main 
concerns of the SUSY search strategists at the LHC.
   
Very heavy sparticles, however, tend to violate the 
naturalness criterion 
\cite{natural1,natural2}. There are attempts to quantify
the degree of 
naturalness violation by defining a set of fine-tuning 
parameters \cite{barbieri}. Large magnitudes of these parameters
increase the degree of violation. 
Thus models with heavy sparticles have been seriously considered  
only if  the magnitudes of the fine-tuning  parameters turn out to be 
small. A case in point is focus point supersymmetry \cite{focus}. In 
this particular parameter space of the minimal
supergravity(mSUGRA) \cite{msugra} model, 
the scalar superpartners turn out to be  very heavy. Nevertheless the 
naturalness parameters have acceptably small magnitudes. 
Moreover, the signatures of 
the gluinos ($\tilde g$), which could be relatively light, are easily 
observable at
the LHC \cite{signatures} if the gluino mass($m_{\tilde g}$) is $\lsim$ 2 TeV. 

However, it has recently been emphasized that SUSY has many other 
attractive features apart from the solution of the naturalness problem.
The simplest versions of the supersymmetric grand unified theories
(SUSYGUTs) allow the  unification of the strong, electromagnetic and 
weak forces \cite{unification}. Moreover, the stable, neutral
and weakly interacting lightest superpartner (LSP) in a R-parity 
conserving
theory is an ideal candidate for explaining the observed dark matter 
relic density in the universe \cite{dm,wmap}. Thus models have been 
constructed  \cite{split} with 
very heavy scalars (squarks and sleptons) which can explain the coupling
constant unification at the grand unified theory (GUT) scale ($M_G$) and 
the 
relic density data \cite{wmap}. As expected, such models predict large values of the 
fine-tuning  parameters \cite{split}. The additional virtues of these 
models with very heavy scalars are highly suppressed flavour 
changing neutral current (FCNC) induced processes and CP 
violating phenomenon \cite{split}.
The LHC signatures of these `unnatural' split SUSY models, which mainly 
arises
from gluino production and decays have been analysed in great 
details \cite{splitlhc}. It should be borne in mind 
that it is  hard to precisely quantify  
the acceptable magnitudes
of the fine-tuning parameters. Moreover, the fact that one has to live with fine tuned 
parameters (e.g., a tiny cosmological constant) in many other cases, 
provides further motivation for these 'unnatural' models.

In order to settle the all important issue of existence or 
non-existence of SUSY experimentally without  any
theoretical bias,  
one  would like to develop a search strategy at the LHC
which could probe the unexplored parameter space   
just beyond the regions excluded by the searches at the Large 
Electron Positron 
collider (LEP) \cite{lepsusy} and the Tevatron collider \cite{cdf,D0,pape} 
and could continue these probes  upto the kinematic reach of LHC.
The best limits on the masses of the sparticles belonging to 
the electroweak sector comes from the LEP experiments \cite{lepsusy}. 
Typically these mass limits  
are $\approx$ 100 GeV
which is approximately the kinematic reach of LEP experiments 
\footnote{If sneutrinos decay invisibly,  the best lower limit on
their mass ( $\approx$ $M_Z/2$)  from the invisible width of  Z
measured at LEP turns out to be much weaker. A proposal for improving
this limit can be found in
\cite{adasesh}.}.
The best limits on the strongly 
interacting squark-gluino sector come
from the searches at the $p \bar p$ collider Tevatron. These 
limits are, 
however, more model dependent.
The CDF and D0 collaborations have been looking for the
sparticles since the dawn of the Tevatron experiments nearly 20 years
ago \cite{pape}. Assuming
that there are five flavours of squarks of L and R type and each has
approximately the same mass as the gluino 
($m_{\tilde q} \approx m_{\tilde g} =\tilde m$),
the CDF collaboration obtained the limit $\tilde m >$
392 GeV. For heavier squarks $m_{\tilde q} =$ 600 GeV, the gluino mass limit
is $m_{\tilde g} >$ 280 GeV \cite{cdf}. The D0 collaboration has obtained similar
limits \cite{D0}.

The above limits imply that in a subspace of the  parameter space yet to be probed
experimentally, the  masses 
of the sparticles belonging to the electroweak sector  
could be well within the reach of the LHC ($\geq$ 100 GeV), while the 
strongly interacting sparticles could be rather heavy (close to the 
kinematic reach 
of the LHC or even beyond it). 
In this paper we focus attention on this parameter space. 
From the point of 
view of  LHC search strategies this scenario is challenging  
since signals from the strongly interacting sparticles 
most easily accessible at hadron colliders are absent.
Henceforth in this paper we shall refer to this scenario
as the light electroweak gaugino-slepton scenario (LEWGSS).

Of course this sparticle spectrum is not realized in conventional 
models of SUSY breaking like the much advertised 
mSUGRA model \cite{msugra}. In the latter model
there are only five free parameters \cite{susy}
namely a common scalar mass ($m_0$), 
a common gaugino mass ($m_{1/2}$), $tan \beta$, $A_0$ and sign of $\mu$.
Here $tan \beta$ is the ratio of the vacuum expectation values of the
two Higgs bosons in the model, $A_0$ is the trilinear soft breaking term
and $\mu$ is the higgisino mass parameter; the magnitude of
$\mu$ is fixed by the radiative electroweak symmetry breaking condition
\cite{ewsb}.
As a result it leads to a highly correlated sparticle mass spectrum
which is rather restrictive.
However, this is certainly  
allowed in more general frameworks and should receive  due 
attention since the mechanism of SUSY breaking is still unknown. 
Later on we shall discuss some  possible 
theoretical frameworks leading to such sparticle spectra
and the impact of the indirect constraints (e.g., the constraints
imposed by  the observed DM relic density \cite{wmap}) on this parameter 
space.

That the SUSY signals in this case may not be easily accessible
even at the LHC
can be anticipated from the simulations by the CMS (see \cite{cms}
 Fig. 13.32 (left)) and the 
ATLAS \cite{atlas} collaborations in the mSUGRA model.
From the scalar sector only the 
dilepton $+ \etslash$ signature from slepton 
pair production may be observable  for slepton mass ($m_{\slep_L}$) 
$\lsim$ 
300 GeV for an integrated luminosity ($\lum$) of 60 $fb^{-1}$
\cite{cmsslep}. However, the existence of this signal alone can 
hardly establish  SUSY.

The 
only other viable signal within the framework of the minimal 
supergravity 
(mSUGRA) model from the electroweak sector alone is the hadronically 
quiet trilepton ($3l$, $l$ = e or $\mu$) events \cite{trilep} from the 
production of the 
lighter chargino ($\chonepm$) and the second lightest neutralino 
($\lsptwo$) pairs. But the chargino-neutralino
mass reach attainable is 
rather modest. Observable signals correspond to m$_\chonepm < $ 107 GeV 
for $\lum$ of 10 $fb^{-1}$ \cite{cms} if the 
slepton mass ($m_{\slep}$) is much larger
( m$_{\slep_R} > 500 GeV$) than the lighter chargino and 
the 
second lightest neutralino. In this case  
the produced gauginos decay via three body modes with branching ratio (BR)s
very similar to that of the W and Z bosons.
As a result the 
produced chargino/neutralino pairs decay into $e$, $\mu$ and $\tau$ 
channels with nearly equal BRs irrespective of the 
specific magnitude of  m$_{\slep}$. These leptonic BRs are relatively 
small since
the hadronic decays of $\chonepm$ and $\lsptwo$ dominate.

However, in a large region of the parameter space of interest sleptons 
could be lighter than the $\chonepm$ and the heavier neutralinos. In this 
case the $\chonepm$ ($\lsptwo$)  decays into two body final states 
involving slepton-neutrino and sneutrino -lepton (slepton-lepton and 
sneutrino-neutrino) pairs of all flavours. Due to the absence of the 
hadronic channels the combined BRs of these leptonic modes 
are nearly 100$\%$  and consequently in a limited region of the parameter 
space the $\chonepm$ 
mass reach can be marginally improved compared to the one presented 
in the last paragraph ( m$_\chonepm < 136$ GeV for m$_{\slep_R} \approx 
100 $ GeV and $\lum$ = 10 fb$^{-1}$) \cite{cms}. 

In either case the mass reach is not 
much larger than the current LEP lower bounds. We also emphasize that
almost the entire parameter space probed in \cite{cms} accessible to the 
clean trilepton signal is 
forbidden by the lower limit on the lightest Higgs boson mass
($m_h$): $m_h > 114.7$ GeV due to the special correlations among the 
sparticle masses in mSUGRA. The only exception is the region with
$m_0 \geq $1400 GeV.     

The absence of viable signals from the electroweak sector is not a 
serious hindrance in the mSUGRA model, since a large region of this
parameter space can, in any case, be scanned via the squark-gluino events. 
However, in a more general framework like the
LEWGSS, with the squarks and the gluinos 
beyond 
the kinematical reach of the LHC,  
the electroweak sector could be the only source of information 
on SUSY. Thus one would like to optimize the search strategy for
this sector.

It should be emphasized that in the parameter space where the leptonic 
two body decays of  
$\chonepm$ and $\lsptwo$ dominate, the lighter stau mass eigenstate 
($\stau_1$) is often
significantly lighter than the other sleptons due to mixing in the 
$\stau$ mass matrix \cite{susy}. Thus the chargino-neutralino pairs may 
preferentially decay into final states involving one or more $\tau$ -jets 
and the $3l$ signal could be degraded or even  depleted for all practical 
purposes. We refer to this parameter space as the '$\tau$ - corridor'.

It may be noted that scenarios with light $\stau$ mostly in the context
of the mSUGRA model have been considered in the literature \cite{baer}.
These studies, however, concentrated on the impact of the
light $\stau$ scenario on the signals from squark-gluino events. 
The direct signals from a relatively light electroweak sector did not 
receive the due attention. 

In this paper we propose that signals with $1l + 2 \tau$-jets, $2l + 1 \tau$-jet 
in the final state be also included in the search strategy 
especially if the strongly interacting sparticles are very heavy. 
These signals will be the main discovery channel in the $\tau$ corridor
and, as shown in a subsequent section, may extend the chargino mass 
reach significantly. Moreover,
these final states in conjunction with the conventional $3l$ signal, will 
cover a 
significantly larger parameter space even outside the corridor, 
where two body decays of 
$\chonepm, \lsptwo$ still dominate and extend the chargino mass reach 
appreciably. 

It is well-known that the chargino-neutralino signals - like all other 
SUSY signals - is accompanied by large missing transverse energy 
($\etslash$). Unfortunately 
the $\etslash$ spectrum accompanying the $3l$ signal turns out to be 
rather soft. Infact it has been already observed in \cite{cms} that 
cuts on $\etslash$ do not improve the signal-background ratio since
the signal is also considerably depleted by this cut.

However, the final states with $\tau$-jets have additional neutrinos
and the $\etslash$ spectrum is much harder as we shall show. Thus a 
suitable cut on 
this variable may discriminate against the SM backgrounds more effectively
and compensate for the reduction in the signal size due to limited 
$\tau$-jet detection efficiency. It has been noted that there are 
hitherto neglected  backgrounds 
(mainly from heavy flavour production) 
to the $3l$ signal which reduces its 
visibility \cite{zack}. In principle these backgrounds may affect the 
signals 
involving $\tau$-jets as well. It is reassuring to note that these new 
backgrounds also involve rather soft $\etslash$ spectrum \cite{zack} and 
may be drastically reduced by a strong cut which does not affect the 
signal very much (see section 2 for the details).

In order to illustrate the proposed signals and the 
expected  improvements in the chargino  mass reach,  
we simulate  $\chonepm-\lsptwo$ production followed 
by their decays using the event generator Pythia (version 6.409) 
\cite{pythia} in section 2. In the same section we also
comment qualitatively on the compatibility of the LEWGSS in some 
specific SUSY breaking models. We also comment on the compatibility
of this model with indirect constraints like the one from the dark 
matter relic density. 

In the next section we revisit the mSUGRA model
with low $m_0$ and $m_{1/2}$ where two body decays of 
$\chonepm$ and $\lsptwo$ dominate. This region is ruled
out by the lower bound on the lighter Higgs scalar mass from LEP only 
if the trilinear SUSY breaking coupling $A_0$ is chosen to be zero.
However,  for non-zero trilinear coupling
this parameter space is compatible with both the Higgs mass bound and 
the WMAP data on dark matter relic density \cite{debottam}. More
importantly  this parameter space  leads to novel signals with
$\tau$-rich final states 
\cite{debottam,nabanita} at the LHC. We shall, 
therefore, analyse the prospect
of the signals with $\tau$-jets proposed in this paper and revisit
the conventional $3l$ signal.
We shall summarise the results and future outlooks in 
section 4.\\       
\section{The signals with and without $\tau$-jets}

In this section we focus our attention on signals involving one
or more $\tau$-jets arising from  $\chonepm$-$\lsptwo$ pair production
at the LHC in the LEWGSS. Throughout this paper all masses and 
parameters which have dimension of mass are expressed in GeV unless
otherwise stated explicitly.

We have generated the sparticle spectrum using SUSPECT (version 2.3) 
\cite{suspect} with 
the following choice of the weak scale parameters: a 
common mass
for the L and R type weak eigenstates of sleptons of all three generations
(m$_{\wt{l_L}} = $ m$_{\wt{l_R}}=$  m$_{\wt{l}}$). 
We shall comment on this slepton spectrum  which is 
somewhat different from the one in the mSUGRA model.
For simplicity we have also assumed $M_1 \approx 0.5 M_2$, 
which is the 
typical expectation in a model with a unified gaugino mass in the 
electroweak sector at a high scale (the GUT scale, say) \cite{susy}. 
However, this is not a crucial 
assumption for the viability of the proposed signals as long as the LSP 
is significantly lighter than the sleptons. We have also assumed  
the $\tilde g$ and L and R squarks belonging to all three generations to be 
very heavy: m$_{\tilde g}$ =  m$_{\tilde q}$ = 3.0 TeV. So far as the 
trilepton signal goes it hardly matters even if the strongly interacting
sparticles are assumed to be even heavier. 

We further assume that $tan \beta = 10$ , 
$A_{0} = 0$, m$_{A} = 1000$, where $m_A$ is the mass of the 
pseudo-scalar Higgs boson. 
The last choice, which leads to a 
decoupled
Higgs sector with only one light neutral standard model like 
Higgs scalar, is also not very crucial for the  signals from
the electroweak gauginos.

The size of the signals of interest crucially depend on the mixing in 
the $\stau$ sector driven by the parameter X$_{\tau} =$ m$_{\tau}$ 
($A_{\tau} - \mu~ tan \beta $). Earlier several authors have addressed 
squark-gluino production in mSUGRA by considering large 
$tan \beta$ \cite{baer} only. We have 
varied the parameter $\mu$ to investigate the effect of this mixing. We 
have represented the large (small) $\stau$ mixing scenario by $\mu = 
$1000 (500). Since the dependence of the chargino-neutralino sector on 
$\mu$ and $tan \beta$ are very different, both approaches seem to be 
worth investigating unless additional theoretical assumptions like the 
common 
scalar mass $m_{0}$ in mSUGRA fixes $\mu$ completely from EW symmetry 
breaking \cite{ewsb}.

Over the entire parameter space scanned by us
m$_{\stau_{1}} \lsim$ m$_{\wt e_{L,R}}$ ,  m$_{\wt \mu_{L,R}}$
due to mixing effects in the $\wt \tau$  mass matrix. The mass difference
of course increases for larger mixing. 
\begin{table}[tb]
\begin{center}
\begin{tabular}{|c|c|c|}
\hline\hline
($m_{\wt l_l}, m_{\lsptwo}$) & (160,274) & (270,274)    \\
\hline
Decay modes &&\\
$\chonep \ra \snu_l l^+            $ & 37.8 & 38.6  \\
$\chonep \ra \snutau \tau^+        $ & 18.9 & 19.4  \\
$\chonep \ra \wt l_l \nu_l         $ & 27.8 & 1.6  \\
$\chonep \ra \stau_1^{+} \nu_{\tau}$ & 10.7 & 20.1 \\
$\chonep \ra \stau_2^{+} \nu_{\tau}$ & 4.2  & -- \\
$\chonep \ra \lspone W             $ & --   & 21.5  \\

\hline \hline
$\lsptwo \ra \wt l_l l             $ & 30.6 &  2.8 \\
$\lsptwo \ra \stau_1 \tau          $ & 11.6 &  22.5 \\
$\lsptwo \ra \stau_2 \tau          $ & 4.8  &    -- \\
$\lsptwo \ra \snu \nu              $ & 52.5 &  57.2 \\
$\lsptwo \ra \lspone h             $ & --   & 17.0   \\
$\lsptwo \ra \lspone Z             $ & --   &  --  \\
\hline \hline
\end{tabular}
\end{center}
\caption{The BRs of the dominant decay modes of the $\chonep$ and $\lsptwo$
for $\mu = 500$ at two representative points inside and outside the
'$\tau$ corridor'.}
\end{table}

In Table 1 we present the BRs of the lighter chargino (see the upper 
half) and the second lightest neutralino (see the lower half) decays 
in two representative  scenarios. In the first case (see the second 
column) the common slepton mass is much smaller than the chargino mass. 
Consequently BRs of
the two body chargino and neutralino decays in  different leptonic 
channels are  approximately the same inspite of the fact that
the $\stau_1$ lighter than the other sleptons.

Through the second scenario we illustrate the '$\tau$ corridor' defined in the 
introduction. Here the $m_{\slep}$ is close to
the chargino mass, yet the $\stau_1$ is considerably lighter
due to mixing. The resulting BRs are presented in the third column  
of Table 1. These suggest that the $3l$ signal will be heavily 
suppressed compared to case 1,
but the final states with $1\tau + 2l$ and $2\tau + 1l$ may produce 
observable signals.

We have simulated all events with gaugino pair production by 
Pythia(version 6409) \cite{pythia} . Initial
and final state radiation, decay, hadronization, fragmentation and
jet formation are implemented following the standard procedures in 
Pythia. The backgrounds have also been simulated by Pythia. 

We next discuss different signals and the
kinematical cuts for improving their size relative to the 
background.

For events with $\tau$-jets, the parent $\tau$s are selected with 
P$\mathrm{_T \ge 20}$ and 
$\vert\eta \vert < 2.4$. The  $\tau$-jets are then
divided into several $E_T$ bins
from 30 to 200.
A $\tau$-jet  in any  bin is then treated as tagged
or untagged according to the efficiency ($ \epsilon_{\tau}$) given in
\cite{cms1} for that bin. Isolated leptons 
$(l=e,\mu)$ 
are
selected if P$^e{\mathrm{_T \ge 17}}$ and
P$^\mu{\mathrm{_T \ge 10}}$ and $\vert\eta^{l} \vert < 2.4$.
For lepton-jet isolation we require $\Delta R(l,j) > 0.5$, where
$\Delta R = \sqrt{(\Delta \eta)^2 +(\Delta \phi)^2 }$.
The detection
efficiencies of e and $\mu$  are assumed to be $ 100 \%$ for 
simplicity. 
 An invariant mass cut on the opposite 
sign dilepton pair 80 $< M_{inv}^{ll} <$ 100 is employed for 
the $1\tau + 2l$ signal to remove the backgrounds from the Z-bosons. We 
have further rejected all  events 
with tagged $b$-jets  to reduce the $t \bar t$ background.
 This reduces the $t \bar t$ background
from 0.064 pb to 0.012 pb. In contrast  the signal, e.g., with $\mu = 
500$, 
$m_\chonepm = 253.7$ and m$_{\slep} = 150$
reduces from 0.0042 pb to .0041 pb.
On the other hand the usual veto against light flavour jets other 
than the tagged $\tau$-jets 
is not used as it does not improve the $S/\sqrt{B}$ ratio, 
where $ S(B)$ is the total number of
signal (background) events.

In order to examine whether the events with $\tau$-jets, if 
combined  with the clean $3l$ events, improve the chargino 
mass reach at the LHC, we have also simulated the later events using 
Pythia.   
However, since the purpose of this generator level work is to
suggest a new possibility for chargino-neutralino search and
not to present a complete analysis, we do not make a full background 
analysis.  We simply follow the analysis of CMS collaboration for 
the $3l$ signal and use 
the total background 
estimated by them (0.05953 pb) (see \cite{cms}, section 13.14) . The 
cuts imposed on the 3$l$ the signal  are summarized below. 
(i)Events with 3 $leptons$ with $P_{T}^{l} >10$, 
$\vert\eta^{l} \vert < 2.4$ are selected. (ii) Events with jets having 
transverse energy E$_{T} 
> 30$ and  $\vert\eta \vert < 2.4$ are vetoed. 
(iii) Events involving two same flavor opposite 
sign (SFOS) isolated leptons (e or $\mu$) in $\vert\eta \vert < 2.4$ 
with 
P$_{T}^e > 10$, P$_{T}^ \mu > 17$ and the dilepton invariant 
mass below the $Z$ peak M$_{ll} < 75$ are retained. The third 
lepton in the event is required to have 
P$_{T}^{\mu,e} > 10$ in $\vert\eta \vert < 2.4$. We note in passing 
that vetoing events with tagged b-jets, as suggested above, 
would further suppress the background 
while leaving the signal practically unaffected.
\begin{figure}[tb]
\begin{center}
\includegraphics[width=\textwidth]{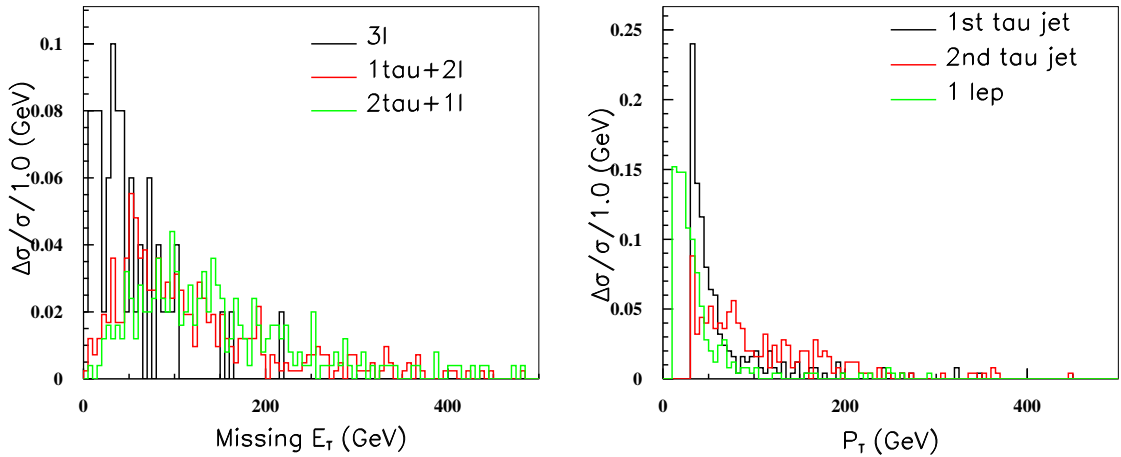}
\end{center}
\caption{ The normalized distribution of missing  transverse
energy ($\etslash$) (left)
after applying all cuts except the one on $\etslash$ and
the ordered transverse momentum (P$_{T}$) distribution of the two
$\tau$-jets  and the lepton
for 2$\tau + 1l$ events (right) after
selection cuts on the 
lepton and the $\tau$-jets.
The details of the parameter space is given in the text.}

\end{figure}


The $\etslash$ spectrum of the $3l$, $1\tau + 2l$ and $2\tau + 1l$ 
events, can be seen in  Fig. 1 (left).  All distributions are
obtained after applying all but the  $\etslash$ cut.  
The plots in Fig. 1 correspond to $\mu = 500$,m $_{\chonepm} = 253$ and
m$_{\slep} = 170$ other parameters are given at the begining of this section.
It is clear from the $\etslash$ plot
that the events containing 
one or more $\tau$-jets have significantly 
harder $\etslash$ distribution than the 
$3l$ events. In fact it has already been noted 
in \cite{cms} and, more recently, in \cite{zack} that the rather soft
$\etslash$ distribution of the $3l$ signal does not permit an  
improvement of its significance by a strong $\etslash$ cut. 
For $2\tau + 1l$ and $1\tau + 2l$ events we apply a 
cut  $\etslash>$100 for background rejection.
We also present the P$_T$ distribution of $2\tau + 1l$ events
in Fig. 1 (right) for the same
spectrum, from where it is clear that the signal
will involve high $P_T$ $\tau$-jets which are taggable
with high efficiencies according to the simulation by 
the CMS collaboration \cite{cms1}.

We have analysed the following backgrounds for the
$\tau$-jet  
signals: WW, WZ, ZZ, $t \bar t$, QCD, $Zb \bar b$, $Z + jets$,
$W \gamma^*/Z^*$. It may be recalled that the importance of the last 
process and its interference with WZ amplitude  
has often been emphasized in the past \cite{zack,others} although 
several analyses
neglected it. We have generated these events using CalcHEP 
\cite{calchep} 
and interfaced them with Pythia. We find that the strong $\etslash$ cut
efficiently removes these backgrounds. The backgrounds involving heavy 
flavour (e.g., $Zb\bar{b}$ events) are also potentially dangerous for 
the $3l$ signal \cite{zack}. They are unimportant for 
the signals involving $\tau$-jets for two reasons.
First, the probability of $\tau$-emission from heavy flavour decay is
significantly smaller than that for e/$\mu$ emission. 
In addition the $\etslash$
spectra of these backgrounds are also rather soft \cite{zack}.

For the $2\tau + 1l$ events the largest surviving background after all 
cuts comes
from  $W \gamma^*/Z^*$ ( 0.000502 pb). For $1\tau + 2l$ signals,  
$t \bar t$ events play this role with a size of  0.012 pb, $W 
\gamma^*/Z^*$ contributes 0.000314 pb.
We summarize the observability of various  signals in Fig. 2 (Fig. 3) for 
the large mixing (small mixing) case. 
Throughout the white region to the left of the blue
dotted line corresponding to $m_{\stau_1} = m_{\lspone}$, 
$\stau_1$ is the LSP and it is theoretically disallowed.
The red line corresponds to $m_{\chonepm} = m_{\slep}$. No observable 
signal can be seen in the remaining white regions in the figures.  
\begin{figure}[tb]
\begin{center}
\includegraphics[angle =270, width=0.8\textwidth]{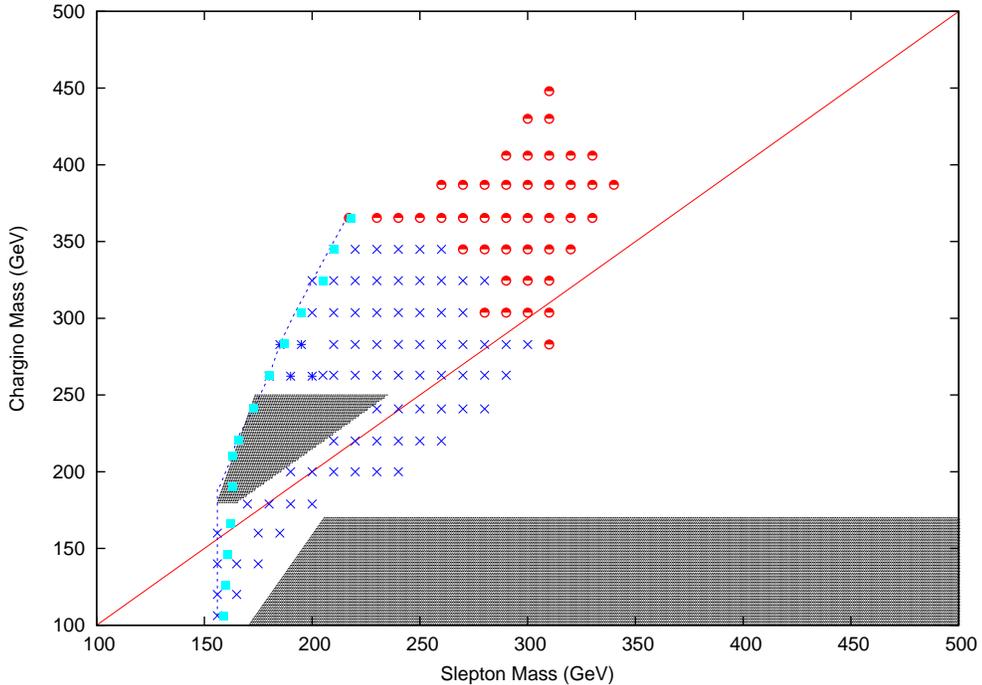}
\end{center}
\caption{Regions of $m_{\chonepm}$ - $m_{\slep}$ plane 
which could be probed by (i)$2\tau + 1l$ (ii)$1\tau + 2l$
and (iii)$3l$ signals for $\mu = 1000$ (large $\stau$ mixing scenario), $A_0 = 0$ and $tan \beta  =10$
(see text for details).}
\end{figure}

\begin{figure}[tb]
\begin{center}
\includegraphics[angle=270, width=0.8\textwidth]{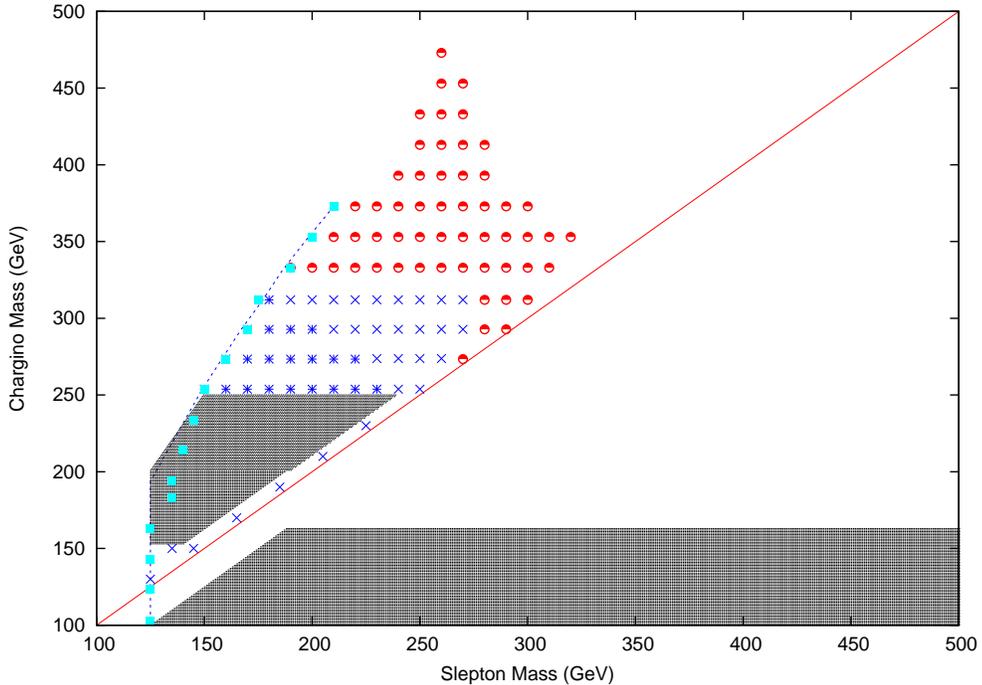}
\end{center}
\caption{Regions of $m_{\chonepm}$ - $m_{\slep}$ plane
which could be probed by (i)$2\tau + 1l$ (ii)$1\tau + 2l$
and (iii)$3l$ signals for $\mu = 500$ (small $\stau$ mixing scenario), $A_0 = 0$ and $tan \beta  =10$
(see text for details).}

\end{figure}


The blue crosses in both the figures indicate the parameter space where
5$\sigma$ signal via the combined 
$1\tau + 2l$ and $2\tau + 1l$ events can be achieved with $\lum = 10 fb^{-1}$.  
Clearly a significantly larger parameter space can be
covered compared to the pure $3l$ signal. In the large mixing case
one can even probe some parameter space where the common slepton mass 
is larger than the m$_{\chonepm}$.  The best chargino mass reach is
estimated to be  m$_{\chonepm} \leq 350$ (330) for 
$m_{\slep} \le$ 250 (290) in the large mixing (small mixing) case. 
The red circles indicate the regions where 3$\sigma$ signal is 
achievable through these channels. For higher 
luminosity the  reach can certainly be extended. For example, with 
m$_{\chonepm}$= 500 (425), m$_{\slep}$ $= 300$, a $5 \sigma$ signal at 
$300 
fb^{-1}$ can be attained  in the large mixing (small mixing) scenario. 

It may be interesting to compare the LHC mass reach with the current
limits obtained from $3l$ events at the Tevatron. The CDF collaboration,
for example, analysed the small $\tau$-mixing scenario in mSUGRA with
tan$\beta =$3 \cite{teva3l}. Analysing 2 fb$^{-1}$ of data they obtain
lower limits of 145 (127) on $\mchonepm$ if two body (three body) decays of
$\chonepm$ and $\lsptwo$ dominate.

Even in the regions where the clean 3$l$ channel alone gives a satisfactory 
signal, the inclusion of the $2\tau +1l$ and $1\tau + 2l$ events
may improve the overall statistical significance.
This could be especially important in view of the observation that
several new backgrounds neglected in the earlier analyses
may reduce the currently estimated
statistical significance of the $3l$ signal \cite{zack}.

The grey shaded regions in Figs. 2 and 3 corresponds to  
$S/\sqrt(B) \ge 5$ for the clean $3l$ signal alone. Since 
our parametrization is somewhat different from the mSUGRA scenario 
our signal size differs from the CMS 
analysis. For example, for  very heavy squarks the 
$\chonepm$ - $\lsptwo$ pair production cross section increases by
15 - 20\% compared to the cross sections in mSUGRA for comparable
chargino-neutralino  masses.
 However qualitatively our conclusion is quite similar to \cite{cms}. 
Moreover, we have explicitly checked that for the mSUGRA 
point LM9 our efficiencies agrees with that given in Table 13.15
of \cite{cms}. We have, however, not included the trigger 
efficiency( $\approx$ 80\%) in our analysis, nor have we taken detector 
effects into account. 
The m$_{\chonepm}$ reach is seen to be 250 for m$_{\slep_{L,R}} 
< $ m$_\chonepm$ (i.e, if the 2 body leptonic decays of the electroweak 
gauginos are the dominant channels). 

If the channels involving $\tau$-jets are combined with the $3l$ we 
obtain 5$\sigma$
signal in a sizable region outside the $\tau$-corridor for both small 
and large $\tau$-mixing( see the regions demarcated by blue stars
in Fig. 2 and Fig. 3). 

For 
larger m$_{\slep}$ the decays of $\chonepm$ and $\lsptwo$  
mediated by W and Z respectively dominate. This reduces the leptonic 
BRs and , consequently, the $\chonepm$ mass 
reach is smaller (m$_{\chonepm} < 170$ (see the grey shaded 
areas in Figs. 2 and 3 for large  m$_{\slep}$ ). It should, 
however, be 
stressed that all grey shaded regions are consistent with the lower 
bound on
the Higgs mass from LEP. This is due to the large radiative 
correction to the Higgs sector by the heavy squarks.

The chargino mass reach in this region cannot be further 
improved by combining the $1~\tau +$ 2$l$ and 2$\tau + l$ 
events with the clean trilepton signal. This is because 
here both the signals are
rather weak due to small BRs as well 
as suppression resulting from $\tau$ detection efficiencies.

Throughout this analysis we have used leading order cross sections
for the signals as well as the backgrounds using CalcHEP.
If next to leading order (NLO) corrections are included
the signal cross section is expected to increase by 1.25 to 1.35
\cite{Beenakker}. As noted above $t \bar t$ events are
the dominant background for the $1\tau + 2l$ signals. 
The NLO cross section for $t \bar t$ production
is 800 pb \cite{ttnlo} which is about a factor of two larger than
the leading order cross section used in this paper.
The significance
$S/\sqrt B$ estimated for this signal
in this paper, therefore, will remain almost
unchanged. The relevant backgrounds for the
other signals ($3l$ and $2\tau + 1l$) come from
pure electroweak processes ($W\gamma^*/Z^*$ etc).
One can therefore conservatively conclude that
the total background is not likely to be enhanced
by a factor larger than two.
The significance of various signals estimated by us
are therefore conservative vis-a-vis the
uncertainties in the production cross sections.
  
Close to the line m$_{\stau_{1}} \approx$ m$_{\lspone}$ in Fig. 2 and Fig. 3
the Dark Matter (DM) 
relic density computed by 
the program micrOMEGAs (version 2.0) \cite{micro} turns out 
to be consistent with the 
Wilkinson 
Microwave Anisotropy Probe (WMAP) data \cite{wmap}: 
$0.09 < \Omega_{CDM}h^2 < 0.13$ where, $\Omega_{CDM}h^2$ is the DM
relic density in units of critical density, $h$ is the Hubble constant.
(see the blue squares). Both bulk 
annihilation and $\stau$ 
- coannihilation contributes dominantly to the relic density. 
In the large
$\tau$-mixing scenario (Fig. 2) observable signals may be obtained
in regions consistent with the WMAP data 
through i) the clean $3l$ events alone ii) events of type i) augmented by 
the events containing $\tau$-jets or iii) purely $\tau$-jet type events.
If $\tau$ mixing is small signal  iii) is somewhat disfavoured.   
The sparticle 
spectrum at a representative point is presented in Table 2. At this 
point an 
observable signal is obtained by combining the $3l$ events with the events 
involving the $\tau$-jets.


\begin{table}[tb]
\begin{center}
\begin{tabular}{|c|c||c|c||c|c|}
\hline \hline
$\lspone         $ & 123.6 & $\wt \tau_1$ & 125.4  & $\wt \nu_{l_L}$ & 136.0  \\
\hline
$\wt \nu_{\tau_L}$ & 136.0 & $\wt l_R   $ & 156.3  & $\wt l_l      $ & 156.8  \\
\hline
$\wt \tau_2      $ & 182.4 & $\chonepm  $ & 253.6  & $\lsptwo      $ & 253.8 \\
\hline
$\lspthree       $ & 518.9 & $\lspfour  $ & 531.7  & $\chtwopm     $ & 531.9  \\
\hline
$\wt b_1         $ & 2888.7 & $\wt q_L   $ & 2890.0  & $\wt q_R      $ & 2890.0 \\
\hline
$\wt b_2         $ & 2893.0 & $\wt t_1   $ & 2923.3  & $\wt g        $ & 2942.9 \\
\hline
$\wt t_2         $ & 2956.5 &  &    &  &  \\

\hline \hline
\end{tabular}
\end{center}
\caption{The Mass spectrum in order of ascending masses
 for a representative point $M_1 =125$, $M_2 = 250$,
 $M_{\tilde q,\tilde g} = 3000$, ${m_{\slep}}= 150$, $\mu =500$,$A = 0$,tan$\beta = 10$ and Omega($\Omega$) $= .0903$.}
\end{table}


We have checked that if $A$, $tan \beta$ and $\mu$ are varied, keeping 
X$_\tau$ fixed, the area of the region consistent with WMAP data
increases. The above variations do not affect the $\chonepm$ and 
$\lsptwo$ decay characteristics drastically. 
Since the size of the signal basically depends on masses of the 
$\chonepm$ and the $\lsptwo$  and their BRs, the mass reach remains more 
or less the same.

It should, however, be noted that a large part of the parameter space 
corresponding to the corridor is  disfavoured by the WMAP 
data. Moreover in 
this region the predictions exceed the  observed relic density.
If signals corresponding to this region are observed, then SUSY can not 
be the origin of  the  observed dark matter relic density. But this
region may still be of interest as far as the LHC signals are concerned. 
For example, a tiny R-parity violating coupling - induced, for example,
by higher dimensional operators - 
would make the neutralino stable at LHC experiments 
but unstable cosmologically. Thus the issue of dark matter could be 
completely decoupled from collider signals. 

Large supersymmetric contributions to FCNC processes (the SUSY flavour
problem) or CP violating processes (the SUSY CP problem)
are potentially dangerous and may lead to strong indirect constraints
on model parameters.
The SUSY flavour problem in the squark sector can be
evaded 
or, at the very least, can be  much softened by the large squark 
gluino masses.
We remind the reader that the signal discussed
in the paper are insensitive to $m_{\tilde q,\tilde g}$
once these parameters are  beyond the kinematic reach of the LHC.
The same is true for
the SUSY CP problem in the squark sector. However, some tuning of
the SUSY CP phases and other parameters
may be needed for accommodating the bounds on the
electric dipole moment (EDM)
of the electron. The $g-2$ of the muon may receive a large SUSY
contribution as we have checked with SUSPECT. For the point shown above
this contribution: 2.86 $\times 10^{-9}$ is, however, acceptable.

The only Achilles' heel of the proposed scenario could be 
the high value of the naturalness 
parameter as we have checked with SUSPECT. However, as already noted in the introduction, 
the main interest is to see the predictions of a 
model which retains all  virtues of SUSY except for the naturalness. 

The scenario understudy may arise naturally in gravity mediated SUSY 
breaking provided the gaugino mass universality 
at the GUT scale is given up. Gaugino masses at the high scale are 
generated by a chiral superfield. If this  superfield is singlet
under the GUT group then an universal gaugino mass ($M_1 = M_2 = M_3 = 
m_{1/2}$) emerge at $M_G$. GUT non-singlet superfields in general leads
to non-universal gaugino masses at $M_G$ \cite{nonuniversal}.    
If $M_3 >> 
M_1$,$ M_2$ at $M_G$, then the squarks  and the gluinos will be 
naturally heavy at $M_G$,
while the sparticles belonging to the electroweak sector may have
smaller masses if the common scalar mass $m_0$ and $m_{1/2} = M_1 =M_2$
at the GUT scale are small.  
The clean $3l$ signal in several scenarios with non-universal gaugino 
masses  has recently been studied in \cite{biswarup}.

However, the desired scenario cannot be accommodated in SUSYGUTs 
based on SU(5) or SO(10) if a single chiral superfield, non-singlet
under the GUT group, generates the gaugino 
masses \cite{nonuniversal}. If on the other hand
a combination of several chiral superfields, transforming 
differently under the GUT group, contributes to this non-universality    
then the desired pattern may emerge in principle. The key point is that 
the chiral superfields belonging to different GUT representations 
contribute to the gaugino masses with different
magnitudes and signs. Thus the  mass pattern may emerge 
if a suitable linear combination of these chiral superfields come into
play. 

 With $M_3 =1200$, $M_1= M_2 =300$                 
at the GUT scale, we obtain the spectrum in Table 3 with features
 similar to the spectrum used in the LEWGSS.
We also obtain $\Omega = 0.10$ with the dominant contribution coming
from bulk annihilation (63 \%) while 33\% comes from coannihilation. 
Observable signals with $\tau$-jets in the final state are predicted. 
Coupling constant unification at $M_G$ ($\sim 10^{16}$) also 
occur naturally. 
\begin{table}[tb]
\begin{center}
\begin{tabular}{|c|c||c|c||c|c|}
\hline \hline
$\lspone         $ & 117.2  & $\wt \tau_1       $ & 132.1    & $\wt l_R
$ & 193.6  \\
\hline
$\wt \nu_{\tau_L}$ & 221.6  & $\wt \nu_{\l_L}   $ & 222.5    & $\chonepm
$ & 225.3  \\
\hline
$\lsptwo         $ & 225.3  & $\wt l_l          $ & 235.7    & $\wt
\tau_2  $ & 272.7 \\
\hline
$\lspthree       $ & 1433.5 & $\lspfour         $ & 1435.4   & $\chtwopm
$ & 1436.1  \\
\hline
$\wt t_1         $ & 1913.1 & $\wt b_1          $ & 2101.9   & $\wt t_2
$ & 2125.9 \\
\hline
$\wt b_2          $ & 2240.9& $\wt q_L          $ & 2245.1   & $\wt q_R
$ & 2248.6   \\
\hline
$\wt g           $ & 2594.7 &                     &          &
& \\
\hline \hline
\end{tabular}
\end{center}
\caption{The Mass spectrum in order of ascending masses
for a representative GUT boundary condition $M_1 = M_2 =
300$,
 $M_3 = 1200$, ${m_{\slep}} = 150$, $m_{0} = 160$, $A_{0} = 0$, tan$\beta
= 10$
 and Omega($\Omega$) $= 0.10$.}
\end{table}

However, in this scenario with fixed boundary conditions at the GUT 
scale, $\mu$
at the weak scale is determined by the EW symmetry breaking condition.
The large $\mu$ obtained in this case tend to make the naturalness 
parameters \cite{barbieri}, computed  by SUSPECT, rather large. 
If the squark and 
gluino masses are lowered to 2 TeV and /or
$m_{H_u}$ and $m_{H_d}$ at $M_{GUT}$ are increased relative to $m_0$ the 
magnitudes of these above parameters tend to reduce.

Intuitively another  natural framework for the LEWGSS scenario seems to be the 
gauge mediated symmetry breaking (GMSB) \cite{gmsb}. Here the 
strongly interacting sparticles with
masses proportional to $\alpha_{s}^{2}$, where $\alpha_s$ is
the strong coupling constant, tend to be heavy. In contrast the 
sparticles in the electroweak sector with masses determined by the 
corresponding couplings, are naturally light. However, one may have to go 
beyond the conventional GMSB scenario with the gravitino as the LSP and
$\lspone$  as the LSP. 
In this scenario  the $\lspone$ decays into
gravitino + photon leading to entirely different collider signals. On 
the other hand  models with a $\stau$ NLSP looks 
promising and may lead to models similar to the one
considered here. We leave the other 
details  as  challenges for the model builders.

\section{ The electroweak gaugino signals in mSUGRA revisited}

As noted earlier almost the entire parameter space corresponding to observable 
$3l$ signal in mSUGRA is disfavoured by the lower bound on $m_h$ from LEP data
(see \cite{cms} Fig. 13.32), if
the trilinear coupling $A_0$ is chosen to be zero. For non-vanishing 
$A_0$ even  small values of $m_0$ and m$_{1/2}$ are allowed by the
above bound \cite{debottam}. It is therefore
worthwhile to check the size of the $3l$ and the $\tau$-jet induced  
signals in the 
modified picture. However, these signals are not the discovery channels
since in any case larger signals can be obtained from the squark-gluino events.

The three benchmark mSUGRA scenarios A, B and C \cite{debottam} have common 
m$_{0} = 120$, $tan \beta = 10$ and $\mu > 0$. The values of 
(m$_{1/2}$, $A_{0}$) are (300.0, -930.0), (350.0, -930.0) and
(500.0, 0.0) for A, B and C 
respectively. The sparticle spectrum and the BRs can be seen from Tables
2 and 5 of \cite{debottam}. SUSY events in scenarios A and B will 
contain more $\tau$ than $e$ or $\mu$ showing 
a strong departure from 'lepton universality' 
\cite{debottam,nabanita}. 
However in scenario C lepton universality is restored. 

The dominant DM relic density producing mechanisms in the above 
scenarios are given in Fig. 1(a)of \cite{debottam}.
Scenario A is characterized by both LSP pair annihilation \cite{bulk} 
and 
LSP-$\stau_{1}$ coannihilation \cite{coann}. In scenario B, $\stau_{1}$ 
coannihilation 
dominates among the relic density producing processes although LSP pair
annihilation plays a significant role. 
In scenario C with A$_{0} = 0$ $\stau_{1}$ coannihilation is the only 
DM producing mechanism.  

Using the cuts introduced in the last section we compute the $2 \tau +1 l$,
 $1\tau + 2l$ and $3l$ in the three scenarios.
In scenarios  A, B  the detection efficiencies of the $\tau$-jets
will be rather low due to  low $\stau$ mass. On the the other hand the 
$3l$ signals will be degraded due to small BRs. In scenario C
the observable signal in any channel cannot be found due to small 
production cross section of the $\chonepm$-$\lsptwo$ pair. The results 
are summarized in Table 4. We have also investigated other parameter
spaces consistent with the relic density data. We find 
that none of the three
signals is at the observable level. Hence an important feature of WMAP 
allowed mSUGRA parameter space for low m$_{0}$ and m$_{1/2}$ is that  
signal from direct EW gaugino production is strongly disfavoured even
if $A_0$ is non-zero.
\begin{table}[!ht]
\begin{center}
\begin{tabular}{|c|c|c|c|}
\hline
& A & B & C\\
$\sigma$ (pb)   & 0.747   & 0.403  & 0.106  \\
\hline \hline

2 $\tau$ + 1 $l$              & 0.000149 & 0.000100 & 0.000120    \\
\hline 

1 $\tau$ + 2 $l$              & 0.000172  & 0.000165 & 0.000670  \\

\hline 
3 $l$                        & 0.000179  & 0.000117 & 0.001181  \\

\hline 
\end{tabular}
\end{center}
\caption{Cross-section times efficiency (in pb) of the three signals after all 
cuts mentioned in the text.}
\end{table}


We next investigate the parameter space not allowed by the DM data but
consistent with the $m_h$ bound from LEP (see Fig. 4 of \cite{debottam}).
The choice of parameter is $A_0 = -700$, $tan \beta =10$ and $\mu >0$.
We find that there is a small region where W, Z mediated three body 
decays of $\chonepm$ and $\lsptwo$ dominate and observable $3l$ signal is 
possible. Some representative parameter spaces are: 
i) $m_{1/2}=200$, $210<m_0<300$ 
ii) $m_{1/2}=220$, $230<m_0<280$ 
iii) $m_{1/2}=230$, $240<m_0<250$.    

It is worthwhile to compare the LEWGSS in section
2 with mSUGRA and identify the features of the sparticle spectrum
responsible for the degradation of chargino-neutralino signals in
mSUGRA with low $m_0$ and $m_{1/2}$. We can identify the following
points. The heavy squarks in
the more general scenario yield a larger $\chonepm$-$\lsptwo$
production cross section compared to mSUGRA for the same $m_{\chonepm}$.
In
mSUGRA with a common scalar mass
$m_0$ at $M_G$, $\stau_R$ turns out to be the
lightest charged slepton at the weak scale due to renormalization
group evolution from $M_G$ to $M_{weak}$.
Consequently  the decay
products
of the lighter mass eigenstate $\stau_1$, which is dominantly
$\stau_R$, are
rather soft. As a result the number of taggable $\tau$-jets 
above a certain $P_T$ threshold, 
in the
final state is small compared to the LEWGSS
with a common slepton mass ( $m_{\stau_L} \approx m_{\stau_R}$) at the
weak scale. It is to be noted that 
in principle the mSUGRA condition $m_{\tilde{l}_R}
= m_{\tilde{l}_L}
= m_0$ may hold  at a higher mass scale higher than $M_G$ ( say, at
the Planck scale $M_P$ ). The evolution between $M_P$ and $M_W$
may induce a  relatively large $m_{\tilde{l}_R}$ at $M_W$. Thus
the spectrum of sleptons at the weak scale may contain a heavier right
slepton relative to mSUGRA. In fact in a SU(5) SUSYGUT the bulk of the
R-type slepton
mass at the weak scale  may come from the $M_P$ - $M_G$ evolution
\cite{polonsky}. Thus viable signals involving $\tau$-jets may
also arise
within the framework of gravity mediated SUSY breaking with a
physically well-motivated variation in the boundary conditions at the
high scale. The R-type slepton mass  may also be enhanced by the D-terms
which arise naturally if the rank of the GUT group is reduced after the
break down of the GUT symmetry \cite{dterm}. For collider
signatures in supergravity models with universal boundary conditions
modified by the $SO(10)$ D-terms see \cite{dsignal}.

\section{Conclusions}

We consider an unconstrained MSSM with very heavy strongly interacting
sparticles (close to the kinematic reach of the LHC
or even beyond it) and relatively light electroweak gauginos and
sleptons. In this scenario, referred to as the LEWGSS, 
the slepton pair production via a Drell-Yan
like mechanism and the clean 3$l$ signal via $\chonepm$ and
$\lsptwo$ pair production are the main SUSY search channels
at the LHC. From the existing simulations (see, e.g., Fig. 13.32 of
\cite{cms} and \cite{cmsslep}) it is, however, expected that these
signals are of modest size.

In a significant region of the parameter space under consideration
the sleptons can be lighter than $\chonepm$ and $\lsptwo$. In this case
the final states from the decays of these gauginos naturally contain
more $\tau$ leptons than electrons and muons (see sections 1 and 2).
As a consequence the 1$\tau$-jet (tagged) + 2$l$ and 2$\tau$
-jets (tagged) + 1$l$ signals from $\chonepm$ or $\lsptwo$ production,
either by themselves or in conjunction
with the usual clean 3$l$ signal, may appreciably  extend the reach of
chargino-neutralino search at the LHC. For the signals involving
$\tau$-jets, it is easier to suppress the SM background because of the
harder $\etslash$ spectrum of the signal (see Fig. 1).
We also veto  events with tagged b-jets for rejecting the $t\bar{t}$
background.
The results are summarized
in Figs. 2 and 3. These results are insensitive to the precise values
of the squark and the gluino masses as long as they are near or beyond
the kinematic reach ($\gsim$ 2.5 TeV) of the LHC.
It also follows that if sleptons are much heavier
than the $\chonepm$ or $\lsptwo$ then the signals involving the $\tau$-jets are
not viable and mass reach via the 3$l$ signal alone remain modest as usual.

In this scenario the SUSY induced FCNC processes or CP violating
processes in the squark sector will be naturally small. Coupling
constant unification at $M_G$ occurs as usual. The WMAP data
on dark matter relic density is satisfied over a small but nontrivial
region of the parameter space. The only Achille's heel of the model is
the large values  of the fine-tuning parameters. However, in many other
fields one has to live with unnatural values of parameters and
one  may accept the  model under consideration in this spirit
which is similar to the philosophy of split SUSY.

The theoretical motivations for the above scenario
are briefly discussed in section 2. Gravity mediated SUSY breaking
with non-universal gaugino masses at $M_G$ with the hierarchy $M_3
>> M_2 \approx M_1$ may generate the mass hierarchy between the strong
sector and electroweak sector considered in this paper.
Scenarios similar to the GMSB model where
the soft breaking masses of the sparticles are proportional to the
corresponding gauge coupling is another possibility.

It is shown in section 3 that in the minimal
supergravity model (mSUGRA), however, the signals involving the
$\tau$-jets are not observable due to the special correlations among the
sparticle masses in mSUGRA. In particular the mass relation $m_{\stau_L}
> m_{\stau_R}$ at the weak scale in mSUGRA makes the decay products
of
the lighter physical state $\stau_1$, which is dominantly $\stau_R$,
rather soft. Consequently the number of taggable $\tau$-jets in the
final state is small compared to the LEWGSS considered
in section 2 with  $m_{\stau_L} \approx m_{\stau R}$ at the weak scale.
As discussed in section 3 the relatively heavy $\stau_R$'s at the weak scale
can be realized in
gravity mediated SUSY breaking with theoretically
well-motivated boundary conditions at $M_G$.
Signals involving $\tau$-jets may, therefore, improve the chargino
neutralino mass reach in these modified scenarios.

{\bf Acknowledgment}:
A. D and N. B acknowledge financial support from  Department of Science and Technology,
Government of India under the project  No. (SR/S2/HEP-18/2003).


 \end{document}